# Frequency-Division Phase Random Optimization for High-Speed Arbitrary Optical Intensity Waveform Monitoring Using Opto-Electronic Finite Impulse Response Filters

**ZHEQING SUN,[1] TAKAHIDE SAKAMOTO,[1]**

[1]*Tokyo Metropolitan University, 6-6 Asahigaoka, Hino, Tokyo, Japan*
**tsaka@tmu.ac.jp*

**We propose and demonstrate a high-bandwidth optical intensity waveform monitoring technique based on frequency-division phase random optimization (FD-PRO) using an opto-electronic finite impulse response (OE-FIR) filter. In this technology, phase random optimization (PRO) enable the estimation of the signal spectral phase. Frequency-division analysis (FDA) combined with PRO saves the iteration required for the optimization, accelerating the signal spectral phase estimation. FDA also facilitates the estimation of the signal spectral amplitude. Using FD-PRO, arbitrary optical intensity waveforms can be easily reconstructed without relying on high-speed digital signal processing. Experimental results reveal that the temporal waveforms are successfully reconstructed at 18-ps resolution.**

High-speed optical waveform monitoring techniques are in demand for use in the state-of-the-art optical fiber communication systems and networks operating at 100 Gb/s or higher [1–5]. Waveform monitoring methods are mostly based on high-speed oscilloscopes, relying on ultrafast digital signal processors (DSPs) clocked at 100 GSa/s or higher [1,2]. These technologies cannot be a practical solution to the waveform monitoring because DSP circuits require extremely high-bandwidth parallel and complex structures for pipeline processing, consuming considerable power. Therefore, low-complexity high-bandwidth waveform monitoring systems that do not rely on ultrafast DSPs or high-speed oscilloscopes are necessary.

In this study, we propose and demonstrate a high-speed, and low-complexity optical intensity waveform monitoring method based on frequency-division phase random optimization (FD-PRO) using opto-electronic finite impulse response (OE-FIR) filters. OE-FIR filters are high-speed adaptive filters based on opto-electronic circuit, employing photodiode and analog-circuit finite impulse response (FIR) filters [6–8]. OE-FIR filters provide effective approaches to cost-effectively equalizing high-bandwidth optical waveforms with less latency and grate scalability [8]. In this study, the OE-FIR filters are applied to high-speed waveform monitoring, with the approach remarkably differing from their typical applications for signal equalization. The intensity waveforms of the optical target signals are reconstructed from the tap coefficients of the OE-FIR filter, which are adaptively optimized and determined using the FD-PRO processing. The waveforms are reconstructed by estimating the signal spectral phase and amplitude of the target signals in the PRO. The adopted frequency division analysis (FDA) accelerates the optimization with fewer iterations. The FD-PRO can be simply implemented on a low-speed DSPs enabling waveform monitoring at 100 Gb/s or higher without relying on high-speed DSPs. We experimentally demonstrate measurements of high-speed optical intensity waveforms using FD-PRO with a parallel-extended OE-FIR filter; optical intensity waveform monitoring is performed at a time resolution of 18 ps.

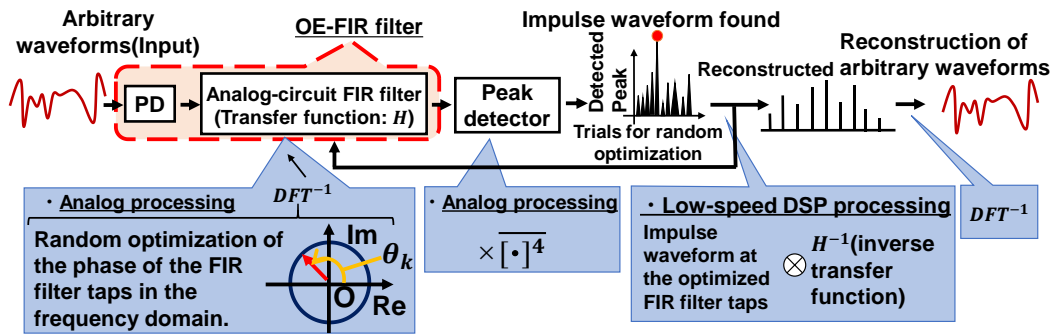

Fig.1 Concept of PRO using OE-FIR for high-speed waveform monitoring

Fig. 1 illustrates the concept of PRO using OE-FIR for high-speed waveform monitoring. The system comprises an OE-FIR filter, a peak detector, and a low-speed DSP. The OE-FIR filter consists of a photodiode (PD) and an analog-circuit FIR filter. The parts of the OE-FIR filter and peak detector are implemented as analog circuits, and the other parts are as low-speed DSP circuits. This configuration facilitates the PRO processing without using high-speed DSPs. The objective of the system is to determine the optimum transfer function that matches the target signal, optimized with multiple iterations controlled by PRO. The signal spectral phases of the target signal are estimated, and its waveforms are accordingly reconstructed. In the PRO processing, the target signal under measurement is input to a high-speed OE-FIR filter that has a transfer function of $H$ tentatively generated in each iteration [5], where $H$ is defined as $H = [h_0, h_1, \ldots, h_{n-1}]$; $h_0, \ldots, h_{n-1}$ are tap coefficients in the OE-FIR. This $H$ is synthesized as the inverse discrete Fourier transfer (DFT) of a complex array, $E = [\xi_0, \xi_1, \ldots, \xi_{n-1}]$, which is called FIR spectral tap coefficients, in this paper. In each iteration in the PRO, the spectral tap coefficients are synthesized in the frequency domain from randomly generated phase set, $\theta_0, \theta_1, \ldots, \theta_{n-1}$, which is related to the spectral tap coefficients as $\theta_k = \arg(\xi_k)$. The equation $H = \mathrm{DFT}^{-1}(E)$ gives us the tentatively generated transfer function $H$. The signal convoluted with $H$ in the OE-FIR filter is input to the peak detector. The detected peak value is recorded and processed using the low-speed DSP. We attempted to maximize the peak value using the iterations, giving different values of $H$ in each time, and the optimal transfer function $H_{\text{best}}$ is saved after identifying the highest peak values. $H_{\text{best}}^{-1}$ provides information on the signal spectral phase of the target signals. Combined with the spectral amplitudes of the target signals identified using the FDA described below, complex spectra of the target signals are identified; accordingly, their waveforms are reconstructed from the complex signal spectra.

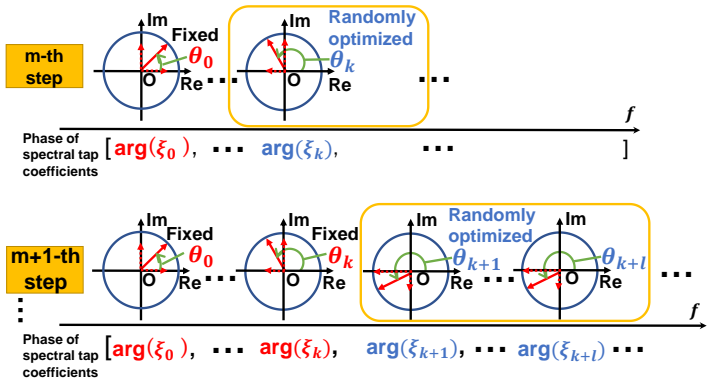

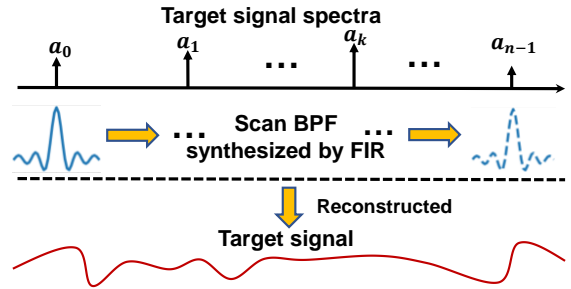

Fig. 3 Signal spectral amplitude estimation by scanning BPF synthesized in OE-FIR.
$a_0, a_1, \ldots, a_k, \ldots a_{n-1}$ are amplitudes of the target signal spectra

Fig. 2 FD-PRO accelerated by FDA.
$\xi_0, \xi_1 \ldots \xi_k$ denote spectral tap coefficients, defined as DFT of tap coefficients assigned to the FIR filters. Their signal spectral phase, $\theta_0, \theta_1, \ldots \theta_k$, are estimated by FD-PRO

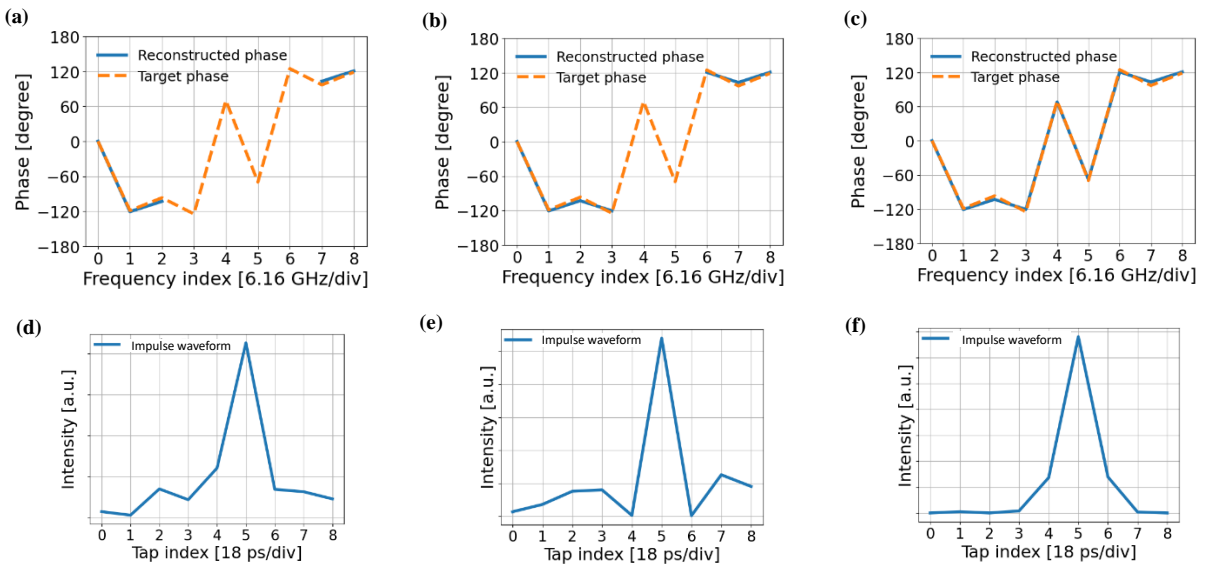

Fig. 4 Numerical results: (a)-(c) signal spectral phases reconstructed by FD-PRO, (d)-(f) impulse waveforms observed when the signal spectral phases have been optimized.

Fig. 2 illustrates the principles of the frequency division analysis (FDA) used for the PRO. The objective is to accelerate the PRO reducing the number of iterations. Initially, the phases of the spectral tap coefficients are optimized in the lower frequency region, before incrementally extending the optimization to the higher frequency region. In the 1st to *m*-th step (*m*: integer) of the iterations, the frequency of the target signal is limited by synthesizing a low-pass filter (LPF) in the OE-FIR filter, and the phase set of the spectral tap coefficients in the limited frequency range, $\theta_0, \ldots \theta_k$, is randomly optimized. Next, in the *m*+1-th step, the cut-off frequency of the LPF is extended, and the phase set of the spectral tap coefficients, $\theta_{k+1}, \ldots, \theta_{k+l}$, where_l: integer, is optimized by inheriting the $\theta_0, \ldots, \theta_k$ optimized in the previous step. This recursive extension of the optimization step, accurately and promptly optimizes all the phases of the spectral tap coefficients, $\theta_0, \theta_1, \ldots, \theta_{n-1}$. As indicated in Fig. 3, the FDA contributes to estimating the spectral amplitudes of the target signals as well. The OE-FIR filter synthesizes the transfer function of the band-pass filter (BPF), and the target signal is spectrally sliced by scanning the center frequency of the BPF passband. Finally, temporal waveforms are reconstructed using the inverse DFT of the complex spectra combining the estimated amplitudes and phases.

Numerical simulations were performed to validate the proposed concept and principles of the PRO. The conditions of the numerical simulation are as follows. In the example shown here, the OE-FIR filter comprised nine taps; the delay of each tap in the FIR filter was 18 ps. The input signal under measurement had a 9-point waveform in each period of 162 ps. The intensity waveform of the target signal in this example had nine samples of [-1.0, 3.0, 6.0, -9.0, 11.0, -6.0, 3.0, 1.0, 5.0] in one period. For the trial in each FD-PRO step, the phase set of the spectral tap coefficients was randomly generated in the range of 0~2π, based on a uniform distribution. The number of FDA division steps in the PRO was 3. A total of 1000 iterations were attempted in each FDA division step. In the first step of the FDA, the cut-off frequency of LPF synthesized in the OE-FIR filter was set to 18.84 GHz and $\theta_0, \theta_1, \theta_2$, the 0th to 2nd spectral tap coefficients within the LPF passband, were optimized and the corresponding signal spectral phases of the target signal was estimated. In the second step, $\theta_0, \theta_1, \theta_2$, optimized in the first step was fixed; the cut-off frequency of the LPF was extended to 24.64 GHz; then, $\theta_3$ was optimized. In the last step, $\theta_4$ was optimized and its corresponding signal spectral phase of the target signal was estimated in the same way.

Fig. 4 depicts the stepwise evolution of the FD-PRO implemented to perform the signal spectral phase estimation of the target signal, in the example. Figs. 4(a)-(c) illustrate the signal spectral phases, estimated at the 1st, 2nd, and the 3rd steps, respectively. Figs. 4(d)-(f) depict the calculated impulse waveforms stored when the highest peak value is found in the FDA division steps. From the optimal spectral tap coefficients obtained when the impulse waveforms with sharp single-peak profiles were found, the signal spectral phases were reconstructed. The reconstructed phase (solid blue) concurs with the phase of the target signal (dashed orange). The results confirm that the spectral tap coefficients of the OE-FIR are optimized and the signal spectral phases [Figs. 4(a)–4(c)] are correctly estimated when the impulse waveform with a sharp peak is clearly detected.

Fig. 5 shows the experimental setup for demonstrating FD-PRO for high-speed arbitrary optical intensity waveform monitoring. In the experiments, we also tested monitoring waveforms of high-speed electrical target signals, for reference. The electrical target signals were generated using an arbitrary waveform generator (AWG) clocked at a sampling rate of 24.6 GSa/s. The target signal had a repetition period of 162 ps; each period of the target signal included nine sampling points. When the optical target signal was generated, a continuous-wave (CW) light was intensity modulated with a Mach-Zehnder modulator (MZM) biased at the quadrature point and driven by the electrical target signal generated from the AWG. The setup for measuring the optical intensity waveforms comprised an OE-FIR filter followed by a peak detector and a low-speed DSP. The OE-FIR and peak detector were in the analog processing part; the rest part is in the digital part. The OE-FIR filter consisted of an PD and a parallel-extended analog-circuit FIR filter. The bandwidth of the PD was 24 GHz. The PD in the OE-FIR was bypassed and the electrical target signal was directly measured when we monitored the electrical waveforms. The parallel-extended OE-FIR filter employed dual analog-circuit FIR filters arranged in parallel. The dual-parallel extension of the OE-FIR filter doubled the time window and the spectral resolution of its filtering characteristics. In the setup, two sets of analog-circuit FIR filters (HMC6545LP5E; Analog Devices, Inc.) were used; each of the FIR filters consisted of nine taps with a time delay of 18 ps between the adjacent taps. A total of 18 points were reserved in each waveform period using the dual-parallel extension. The photodetected signal in the OE-FIR was split in two with a power divider and input to the analog-circuit FIR filters (labeled #1 and #2); the signals output from the FIRs were digitized using a two-channel real-time digital sampling oscilloscope (MSO71254; Tektronix) clocked at 100 GSa/s. The oscilloscope was used to emulate a power combiner used in the OE-FIR to combine the signals output from the FIRs. When the two signals were combined, a 162-ps delay was inserted on one side of the signals to properly extend the time window of the OE-FIR filter. It also emulated the peak detector, where the peak was detected by processing the fourth power and taking its average [5]; the output from the peak detector was recorded using a low-speed analog-to-digital converter (ADC). Using the peak values recorded in the iterations, the FD-PRO implemented in the low-speed DSP estimated the signal spectral phase of the target signal. The number of FDA division steps was 3. In the first step, phases of three spectral tap coefficients around the direct current (DC) frequency were simultaneously optimized using 400 iterations. Subsequently, PRO was performed for each frequency with 50 optimization iterations in each step.

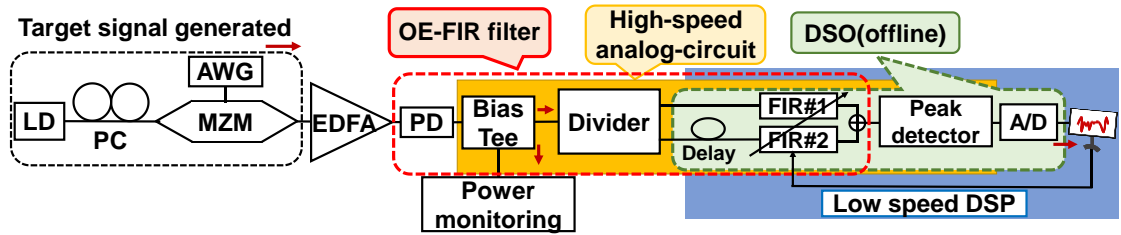

Fig.5 Experimental setup; LD: laser diode, PC: polarization controller, AWG: arbitrary waveform generator, PD: photodiode, OE-FIR filter: opto-electronic finite impulse response filter, FIR: finite impulse response filter, A/D analog-to-digital converter, PC: personal computer, DSP: digital signal processor, DSO: digital sampling oscilloscope

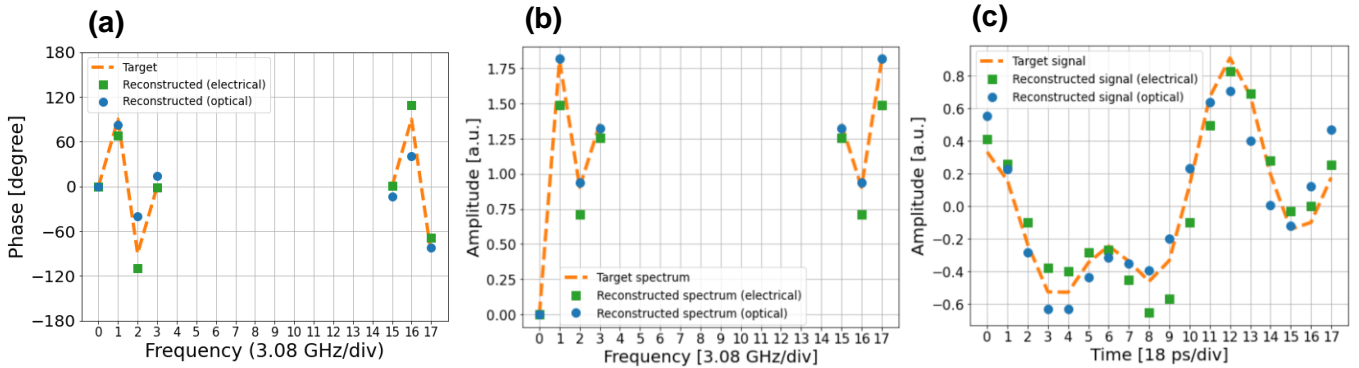

Fig.6 Experimental results; Reconstructed (a) signal spectral phases, (b) signal spectral amplitudes, (c) temporal waveforms

The spectral amplitude of the signal was identified by scanning the BPF synthesized in the parallel-extended FIR circuit. The bandwidth and frequency scanning step of the BPF were 6.16 GHz and 3.08 GHz, respectively. Finally, the optical intensity waveforms (and electrical waveforms for reference) were reconstructed from the estimated signal spectral amplitudes and phases of the target signal. Fig. 6 depicts the experimental results; Figs. 6(a), (b), and (c) indicate the reconstructed signal spectral phases, amplitudes, and temporal waveforms, respectively. In each plot, the dash orange line represents the target spectra and waveforms; the green squares and blue dots in each figure indicate the reconstructed electrical and optical spectra/waveforms, respectively, which concur with the target spectra/waveforms. This indicates that the signal spectral phase and amplitude are accurately estimated using FD-PRO, successfully enabling reconstruction of high-speed optical intensity waveforms.

The demonstration focused on measuring fixed-pattern waveforms. However, the technology is applicable to capture and reconstruct the fundamental waveform of randomly data-modulated signals, such as pulse amplitude modulation (PAM) signals; moreover, combined with coherent detection, quadrature amplitude modulation (QAM) and quadrature phase shift keying (QPSK) signals, which holds promise for waveform monitoring in future high-speed fiber-optic networks.

In this paper, we have proposed and experimentally demonstrated FD-PRO to monitor arbitrary optical intensity waveforms using OE-FIR filters. The signal spectral phase and amplitude of the high-bandwidth optical waveform were estimated, and the temporal waveforms were successfully reconstructed at the time resolution of 18 ps without relying on high-speed DSPs.

**Funding.**
This work was partly supported by CREST, the Japan Science and Technology Agency, Grant Number JPMJCR2103, Japan.

**Acknowledgements**
The authors would like to thank Mr. Shuhei Otsuka for his experimental support.

**Disclosures.**
The authors declare no conflict of interest.